\documentclass[floatfix,aps,pra,twocolumn,showpacs,superscriptaddress,10pt,nofootinbib]{revtex4-2}

\usepackage{graphicx}
\usepackage{dcolumn}
\usepackage{bm}
\usepackage{booktabs}

\usepackage{float}
\usepackage{hyperref}
\usepackage{ragged2e}
\usepackage{amsmath}
\usepackage{xcolor}

\begin{document}

\title{Inverse Physics-informed neural networks procedure for detecting noise in open quantum systems}

\author{Gubio G. de Lima}
\email{gubio.gomes@ufscar.br}
\affiliation{Physics Department, Federal University of São Carlos, 13565-905, São Carlos, SP, Brazil}

\author{Iann Cunha}
\affiliation{Hospital Israelita Albert Einstein, 05652-900, São Paulo, SP, Brazil}

\author{Leonardo K. Castelano}
\affiliation{Physics Department, Federal University of São Carlos, 13565-905, São Carlos, SP, Brazil}

\begin{abstract}
Accurate characterization of quantum systems is essential for the development of quantum technologies, particularly in the noisy intermediate-scale quantum (NISQ) era. While traditional methods for Hamiltonian learning and noise characterization often require extensive measurements and scale poorly with system size, machine learning approaches offer promising alternatives. In this work, we extend the inverse physics-informed neural network (referred to as PINNverse) framework to open quantum systems governed by Lindblad master equations. By incorporating both coherent and dissipative dynamics into the neural network training, our method enables simultaneous identification of Hamiltonian parameters and decay rates from noisy experimental data. We demonstrate the effectiveness and robustness of the approach through numerical simulations of two-qubit open systems. Our results show that PINNverse provides a scalable and noise-resilient framework for quantum system identification, with potential applications in quantum control and error mitigation.
\end{abstract}

\maketitle

\section{Introduction}

The rapid progress in quantum technologies has intensified the need for accurate characterization of quantum systems, particularly in the context of quantum computing and quantum information processing~\cite{wittler2021integrated}. Central to this task is the determination of the system's Hamiltonian, which governs its dynamics and underpins the execution of quantum algorithms. However, realistic quantum devices are inherently open systems, subject to interactions with their environments~\cite{rost2025long}. These interactions lead to decoherence, requiring a more comprehensive description of the dynamics via quantum master equations~\cite{Erhard2019}.

While several traditional techniques exist for Hamiltonian and noise characterization, such as quantum process tomography and Bayesian estimation, they often suffer from high computational cost, demand extensive experimental resources, and are limited in scalability~\cite{doi:10.1080/09500349708231894,PhysRevLett.90.193601,PhysRevA.87.062119}. Recently, machine learning approaches have emerged as powerful alternatives for learning and inferring physical models from data, offering both efficiency and flexibility~\cite{Gebhart2023,fioroni2025learning,schorling2025meta}. In particular, physics-informed neural networks (PINNs)~\cite{RAISSI2019686,doi:10.1126/sciadv.abi8605,doi:10.1137/19M1274067} have attracted significant attention for embedding physical laws, expressed as differential equations, into the training process of neural networks.

In a previous work~\cite{PhysRevA.110.032607}, it was demonstrated that PINNverse can be effectively employed to learn the parameters of a two-qubit Hamiltonian, significantly reducing the number of required experimental measurements. By encoding the system's dynamics through the Heisenberg equation and incorporating data at selected collocation points, it was able to reconstruct the underlying Hamiltonian with high accuracy, even in the presence of noise in the measurements. In this context, collocation points refer to the discrete, equally spaced time steps at which the observables are measured.
In the present work, we extend this framework to open quantum systems governed by Lindblad master equations. This extension enables the procedure to account not only for coherent dynamics but also for environmental decoherence. The resulting architecture simultaneously learns the Hamiltonian and the decoherence rates directly from experimental data. 
Notably, this formulation does not require a huge amount of labeled data to train the machine learning model because this approach is a unsupervised technique that relies only on the expectation values of observables as input data.

Another challenge in quantum computing relies on eliminating crosstalk errors~\cite{Sarovar2020detectingcrosstalk}, which can significantly degrade the performance of quantum gates and measurement processes \cite{gambetta2012characterization, mckay2019three}. Crosstalk  refers to correlations between qubits that should be uncoupled and it arises from imperfect isolation in the control architecture and residual coupling between qubits. In superconducting qubits, for example, crosstalk can result from shared control lines, parasitic capacitance, or the finite anharmonicity of transmon devices, leading to coherent errors such as off-resonant driving and frequency crowding \cite{PhysRevLett.113.220502}. Similarly, in trapped-ion systems, laser beam misalignment and off-resonant excitations can induce crosstalk that affects gate fidelity \cite{PhysRevLett.129.240504}. These effects complicate system calibration and necessitate the development of crosstalk-aware compilation strategies, pulse shaping, and hardware-level mitigation techniques \cite{sundaresan2020reducing,PRXQuantum.1.020318}. Accurate modeling and compensation of crosstalk are essential for scalable quantum computation, particularly in NISQ devices. Here, we also investigate the use of PINNverse as a tool to detect crosstalk errors, in the sense of detecting unwanted interaction between qubits that can be probed through experimental data.

The efficiency and robustness of PINNverse are probed by applying it to representative open quantum systems that exhibit dephasing and relaxation. We further explore the dependence of the accuracy on the number of collocation points and demonstrate the resilience of the method against realistic noisy data. Our findings highlight the potential of PINNverse as a tool for quantum system identification, particularly suited for NISQ devices where noise is unavoidable. 

The remainder of this paper is organized as follows. In Section~\ref{sec:model_methods}, we present the theoretical framework, including the Lindblad equation, the neural network architecture, and the inverse learning scheme. Section~\ref{sec:results} provides numerical and experimental results. Finally, in Section~\ref{sec:conclusions}, we summarize our findings and discuss future directions.

\section{ THEORETICAL MODEL AND METHODS}\label{sec:model_methods}
This work builds upon the methodology presented in Ref.~\cite{PhysRevA.110.032607}, which integrates the PINN framework with a freezing mechanism to achieve efficient learning of Hamiltonian parameters. The PINN procedure enables the training of neural networks by directly embedding the governing differential equations of the physical system under investigation~\cite{RAISSI2019686,doi:10.1126/sciadv.abi8605,doi:10.1137/19M1274067}. This approach offers a significant advantage in reducing the reliance on extensive labeled datasets, as the neural network's learning process is constrained by the inherent physical laws encoded within the differential equations~\cite{RAISSI2019686,doi:10.1126/sciadv.abi8605,doi:10.1137/19M1274067}. Furthermore, we introduce the term \textit{PINNverse} (often referred to as inverse-PINN in the literature) to describe the framework to extract physical parameters from a collection of data by simultaneously satisfying the governing equations and matching experimental observations~\cite{RAISSI2019686,doi:10.1126/sciadv.abi8605,doi:10.1137/19M1274067}. In this context, experimental data is incorporated alongside the differential equations, allowing the model to infer underlying physical parameters.

When dealing to a realistic quantum system, the inherent susceptibility to environmental influences necessitates a departure from idealized closed-system description leading to an open quantum system formulation. Several factors can significantly impact the system's nature, rendering complete isolation from the external environment practically infeasible.
Therefore, the system of interest for quantum computing exhibits decoherence and relaxation due to environmental interactions that imposes critical limitations on the fidelity and scalability of quantum information processing technologies.

The temporal evolution of an open quantum system is generally non-unitary and can be described by a Markovian master equation that governs the dynamics of the system's reduced density operator, denoted by $\rho(t)$. The Lindblad master equation represents a widely employed formalism for this purpose, expressed as:
$$
\frac{d\rho(t)}{dt} = \frac{i}{\hbar} [\rho(t),H] + \sum_k \gamma_k \left( \mathcal{L}_k \rho(t) \mathcal{L}_k^\dagger - \frac{1}{2} \{ \mathcal{L}_k^\dagger \mathcal{L}_k, \rho(t) \} \right),
$$
where $H$ represents the system's Hamiltonian, $\hbar$ is the reduced Planck constant, $\gamma_k$ denotes the decay rate associated with the {\it kth} Lindblad operator $\mathcal{L}_k$, and $\{A, B\} = AB + BA$. The first term on the right-hand side of the Lindblad master equation describes the unitary evolution of the system driven by its Hamiltonian. The second term encapsulates the non-unitary effects arising from the system's interaction with the environment, leading to decoherence. This equation serves as a cornerstone for modeling dephasing and relaxation phenomena. 

To illustrate the aforementioned numerical procedures, we consider a concrete physical realization: a two-qubit system. This choice provides a tractable yet sufficiently complex framework for learning both the Hamiltonian parameters and the decay rates governing the open quantum dynamics. The two-qubit system is also the smallest unity to obtain interactions between qubits and a general learning of larger system can be obtained by the freezing mechanism~\cite{PhysRevA.110.032607}, where external fields are used to isolate a two-qubit system from the larger one. The general form of the two-qubit Hamiltonian can be expressed as a linear combination of tensor products of Pauli matrices:
\begin{equation}
    H = \hbar\sum_{\mu,\nu=0}^{3} J_{\mu,\nu} \, S_{\mu,\nu},
    \label{eqTwoqubit}
\end{equation}
where $S_{\mu,\nu} = \sigma_\mu \otimes \sigma_\nu$, and $J_{\mu,\nu}$ are real-valued coefficients, with $\mu,\nu \in \{0, 3\}$, that determine the contribution of each interaction term.  In Eq.~\eqref{eqTwoqubit}, there are sixteen $J_{\mu,\nu}$ terms characterizing the frequencies and couplings within the two-qubit system. Without loss of generality, the term $J_{0,0}$ is set to zero, as it corresponds to an overall energy shift. Experimental data can be obtained through the expectation value of observable operators, defined as:
\begin{equation}
    \langle S_{\mu,\nu} \rangle(t) = \mathrm{Tr}[\rho(t) S_{\mu,\nu}],
    \label{eq:measurement_trace}
\end{equation}
where $\rho(t)$ is the density matrix of the two-qubit system at time $t$.

The temporal dynamics of these observables are incorporated into the PINNverse framework through the following equation in the Heisenberg picture:
\begin{equation}
    \frac{d\langle S_{\mu,\nu}\rangle(t)}{dt}  = -\frac{i}{\hbar} \langle [S_{\mu,\nu}, H] \rangle(t) +D[S_{\mu,\nu}](t),
    \label{eq:3}
\end{equation}
where the dissipator $D[S_{\mu,\nu}](t)$ is given by:
\begin{equation}
    D[S_{\mu,\nu}](t)=\sum_k \gamma_k \left( \langle \mathcal{L}_k^\dagger S_{\mu,\nu} \mathcal{L}_k \rangle(t) - \frac{1}{2} \langle \{ S_{\mu,\nu}, \mathcal{L}_k^\dagger \mathcal{L}_k \} \rangle(t) \right).
\end{equation}

Here, we assume that the two-qubit system follows a Markovian master equation with dephasing and/or amplitude-damping noise, which are standard model of decoherence channels~\cite{Fonseca_Romero_2012}. The amplitude-damping channel at zero temperature is characterized by the lowering operator $\sigma_{-} = (\sigma_{1} - i\sigma_{2})/2$, while the dephasing channel is represented by the $\sigma_{3}$ operator. This leads to four Lindblad operators acting on each qubit:
\begin{align*}
    \mathcal{L}_1 &= \sigma_{-} \otimes \sigma_{0}, &
    \mathcal{L}_2 &= \sigma_{3} \otimes \sigma_{0}, \\
    \mathcal{L}_3 &= \sigma_{0} \otimes \sigma_{-}, &
    \mathcal{L}_4 &= \sigma_{0} \otimes \sigma_{3}.
\end{align*}
The differential equation \eqref{eq:3} is implemented for all fifteen independent physical observables corresponding to $\langle S_{\mu,\nu}\rangle(t)$ where $\mu$ and $\nu$, without the identity operator. Thus, we get a system of fifteen coupled ordinary differential equations that must be solved given a specific initial condition, which in this case is set to the separable state $|+\rangle^{(1)}|+\rangle^{(2)}$, where $|+\rangle=(|0\rangle+|1\rangle)/\sqrt{2}$.

The central idea of the PINNverse approach is to approximate the solutions of these differential equations using a neural network and to minimize a composite loss function, defined as:
\begin{equation}
    L=L_{m}+ L_{d}.\label{eq:Loss}
\end{equation}
The model-based loss term is:
\begin{equation}
    L_{m}=\sum_{j=1}^{N_t}\sum_{\mu,\nu}\left|\langle \dot{S_{\mu,\nu}} \rangle(t_j)+ i\langle[S_{\mu,\nu},H]\rangle(t_j)-D[S_{\mu,\nu}](t_j)\right|^2,\label{eq:Lmodel}
\end{equation}
where $\langle \dot{S_{\mu,\nu}} \rangle(t_j)=\frac{d\langle S_{\mu,\nu} \rangle(t_j)}{dt}$ and $N_t$ is the number of points used to numerically perform the time evolution of Eq.~\eqref{eq:3} within the PINNverse procedure.
The loss function $L_{m}$ evaluates the dynamical equations at a set of points $\{t_j\}_{j=1}^{N_t}$ in the time domain and the minimization of this term ensures the neural network’s output follows the physical laws imposed by the differential equations Eq.~\eqref{eq:3}. On the other hand, the data-driven loss term is given by:
\begin{equation}
    L_{d}=\sum_{\mu,\nu}\sum_{n=1}^{N_{c}}\left|\langle S_{\mu,\nu} \rangle^{exp}(t_n)- \langle S_{\mu,\nu}\rangle(t_n)\right|^2,
\end{equation}
where $\langle S_{\mu,\nu} \rangle^{exp}(t_n)$ represents the experimentally measured expectation value of the observable $S_{\mu,\nu}$ at time $t_n$, considering $N_c$ collocation points obtained from the experimental data. The loss term $L_{d}$ enforces the constraint that the neural network's predictions align with the experimental observations at the specified collocation points. In this manner, the PINNverse framework compels the neural network to simultaneously solve the differential equations and represent the experimental data. To satisfy both these requirements, the Hamiltonian parameters and the decay rates are consequently learned by minimizing the total loss function.

\section{Results}\label{sec:results}
In this section, we demonstrate the application and robustness of PINNverse with or without the inclusion of an extra noise in the measured data. We begin by considering a two-qubit system, where synthetic data is generated through numerical integration of Eq.~\eqref{eq:3}, which is computed using the \texttt{qutip} library considering the total Hamiltonian given by Eq.~\eqref{eqTwoqubit}. The parameters \( J_{\mu,\nu} \) and \( \gamma_k \)  are randomly sampled from a uniform distribution in the interval $[-\omega_0,\omega_0]$ and $[0,\omega_0]$, respectively. Here, $\omega_0=2\pi/T$ and $T$ is the final time of evolution. 

Subsequently, we explore the effects caused by random error drawn from a Gaussian distribution in the observables for each instant of time. Lastly, we apply the procedure to real experimental data reported in Ref.~\cite{Ficheux2018}, which is relative to a single-qubit system. This step allows us to benchmark the technique under real-world conditions, complementing the simulated scenarios.

\subsection{General two-qubit Hamiltonian}
The first problem we investigate is the most general form of the Hamiltonian for a two-qubit system with all $J_{\mu,\nu}$ terms non-null and considering the amplitude-damping and dephasing channels of noise for each qubit described by the four Lindblad operators $\mathcal{L}_1$, $\mathcal{L}_2$, $\mathcal{L}_3$, and $\mathcal{L}_4$. The accuracy of the parameters estimation is measured through the mean absolute percentage error, which is defined as follows:
\begin{equation}
	\text{MAPE}=\frac{1}{D}\sum_{i=1}^D\frac{|P^\text{exact}_i-P^\text{pred}_i|}{|P^\text{exact}_i|},
\end{equation}
where $D$ is the number of parameters randomly sorted, $P^\text{exact}_i$  ($P^\text{pred}_i$)  denotes the {\it ith} exact (predicted) physical parameter. 

\begin{figure}[!ht]
    \includegraphics[width=1\linewidth]{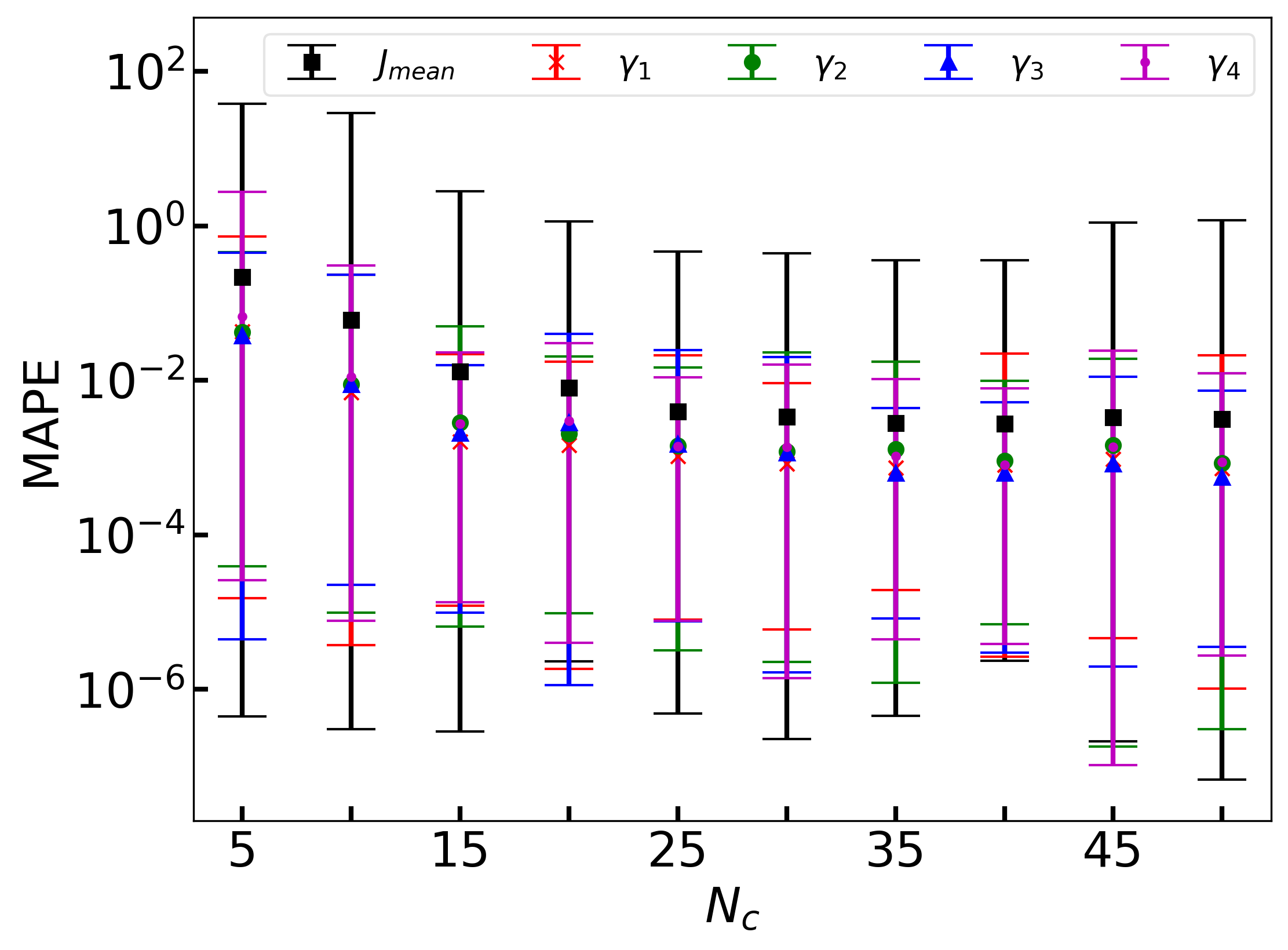}
    \caption{\justifying The MAPE for coefficients $J_{mean}=\sum_{\nu,\mu}J_{\mu,\nu}/15$ and $\gamma_i$ as a function of the number of collocation points $N_{c}$. Symbols indicate the mean values of MAPE, while the error bars represent the maximum and minimum estimated values for each coefficient.}
    \label{fig:Mape_vs_Ndata}
\end{figure}

In Fig.~\ref{fig:Mape_vs_Ndata} we investigate the relationship between the estimation of the  decay rates \(\gamma_i\) and the values of $J_{\mu,\nu}$, as a function of the number of collocation points \(N_{c}\). Here, the values of $\langle S_{\mu,\nu} \rangle^{exp}(t_n)$ are obtained by numerically solving Eq.~\eqref{eqTwoqubit} for randomly sorted values of \(\gamma_i\) and $J_{\mu,\nu}$. These results are plugged into the Loss function defined in Eq.~\eqref{eq:Loss} and our goal is to obtain the randomly sorted parameters previously used to generate the data. The error bars in Fig.~\ref{fig:Mape_vs_Ndata} represent the maximum and minimum estimated values of the various realizations, considering different parameters in each realization of the PINNverse.

In Fig.~\ref{fig:Mape_vs_Ndata} we only provide the MAPE for the mean value $ J_{\text{mean}} =\sum_{\nu,\mu}J_{\mu,\nu}/15$ because there are too many terms that are difficult to distinguish in the same figure. However, the result for each value of $J_{\mu,\nu}$ follows the trend of the mean value result. In Fig.~\ref{fig:Mape_vs_Ndata}, one can see that the MAPE decreases as \(N_{c}\) increases, indicating an improved PINNverse performance when more experimental information is provided. 
For values of $N_{c}$ exceeding 15, the MAPE reaches approximately $10^{-2}$, with a progressively slower rate of decrease as $N_{c}$ increases. This corresponds to an error of approximately $1\%$ between predicted and exact parameters. This low error rate indicates that an accurate estimation can be achieved even with a relative small amount of experimental data.

\begin{figure}[!ht]
    \centering
    \includegraphics[width=1\linewidth]{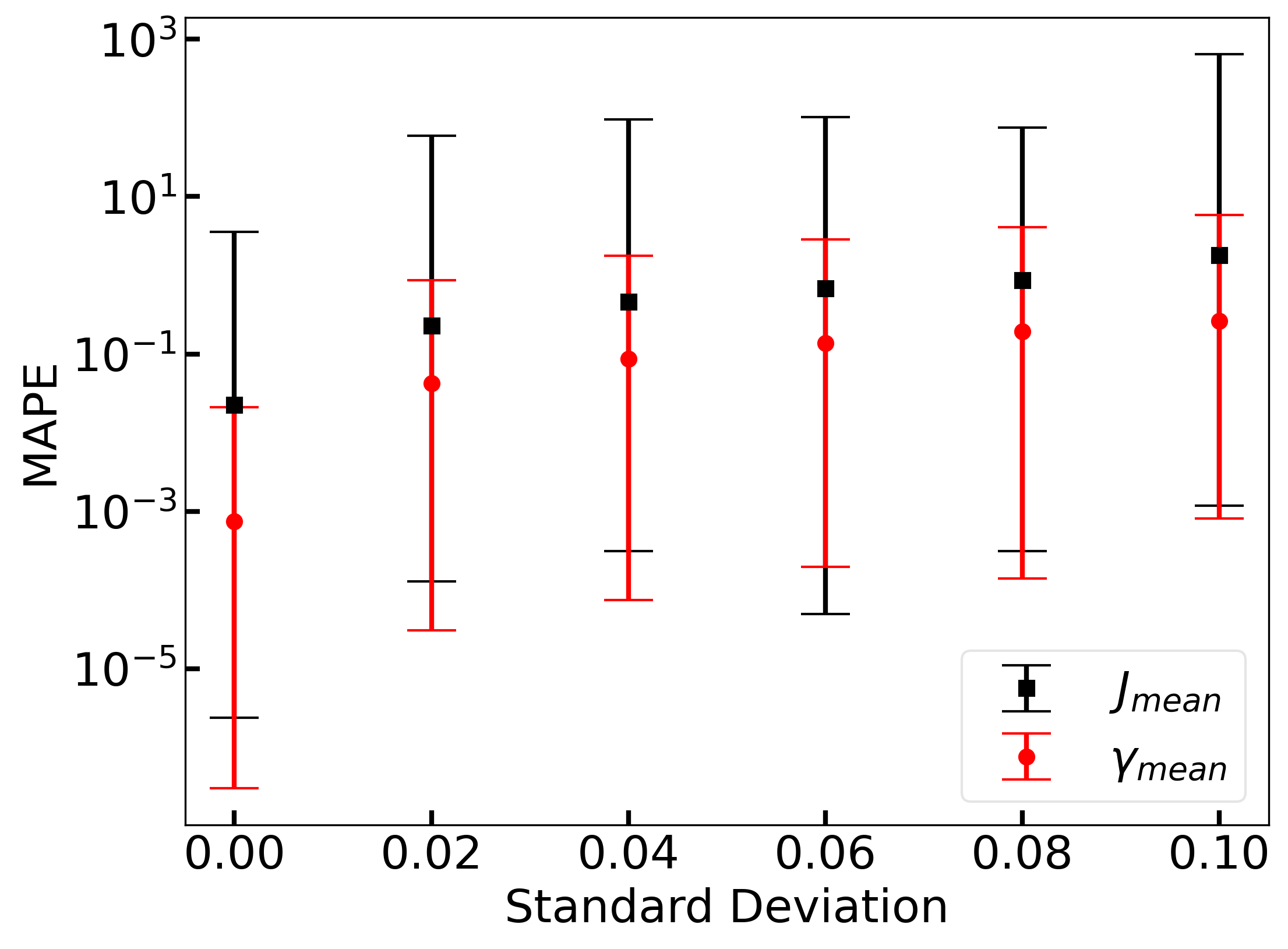}
    \caption{\justifying
    The MAPE for \( \gamma_{\text{mean}} \)(red circles) and coefficients \( J_{\text{mean}} \) (black squares), as a function of the standard deviation of Gaussian noise added to the experimental data considering $N_c$=50. Error bars represent the maximum and minimum estimated values of the MAPE across multiple PINNverse training runs.
    }
    \label{fig:Mape_sdt}
\end{figure}

In Fig.~\ref{fig:Mape_sdt}, we explore the effects of considering a random error in the experimental data drawn from a Gaussian distribution. This type of error represents errors during preparation of the initial state and/or errors in the measurement process. Here, each observable data for $\langle S_{\mu,\nu} \rangle^{exp}(t_n)$ is modified according to $\langle S_{\mu,\nu} \rangle^{exp}(t_n)\rightarrow \langle S_{\mu,\nu} \rangle^{exp}(t_n) + N(0,\sigma^2)$, where $N(0,\sigma^2)$ is the normal distribution with zero mean and standard deviation $\sigma$. Figure~\ref{fig:Mape_sdt} illustrates the sensitivity of the parameter estimation performance of the PINNverse to the level of noise added to the simulated experimental data. Specifically, we report the MAPE for the mean dissipation coefficient \( \gamma_{\text{mean}}= \sum_{k}\gamma_{4}/4\) and the Hamiltonian coupling \( J_{\text{mean}} \), as a function of the standard deviation of Gaussian noise. The results are averaged over multiple independent PINN runs, and the error bars represent the maximum and minimum values of MAPE across these runs.

In Fig.~\ref{fig:Mape_sdt}, one can see that both parameters exhibit an increasing trend in MAPE with rising the standard deviation $\sigma$, reflecting the expected degradation in inference accuracy due to reduced data quality. For noise-free data (\( \sigma = 0.00 \)), the MAPE for \( \gamma_{\text{mean}} \) is on the order of \( 7 \times10^{-4}\)($0.07\%$), whereas for \( J_{\text{mean}} \), it is below \( 0.02 \)($2\%$).  As the standard deviation increases to \( \sigma = 0.02 \), the MAPE for \( J_{\text{mean}} \) reaches values close to approximately \(  2.3 \times10^{-1}\)($23\%$), while \( \gamma_{\text{mean}} \) remains below $4\%$, demonstrating greater robustness of the decay rates estimation under low noisy conditions. These results underscore the critical importance of noise control and data preprocessing when using PINNs for parameter estimation in open quantum systems. While dissipation-related coefficients such as \( \gamma_{mean} \) can be reliably estimated even with moderate noise levels, Hamiltonian parameters require careful treatment to ensure consistent and accurate recovery.

\subsection{Crosstalk errors} 
Crosstalk errors are unwanted correlations between qubits that arises from imperfect isolation of parts in the platform architecture. To understand and mitigate this kind of errors, it is mandatory the identification  
of interaction terms between the two qubits in the Hamiltonian of Eq.~\eqref{eqTwoqubit} where $\mu$ and $\nu$ are different from zero.
\begin{figure}[ht]
    \centering
    \includegraphics[width=1\linewidth]{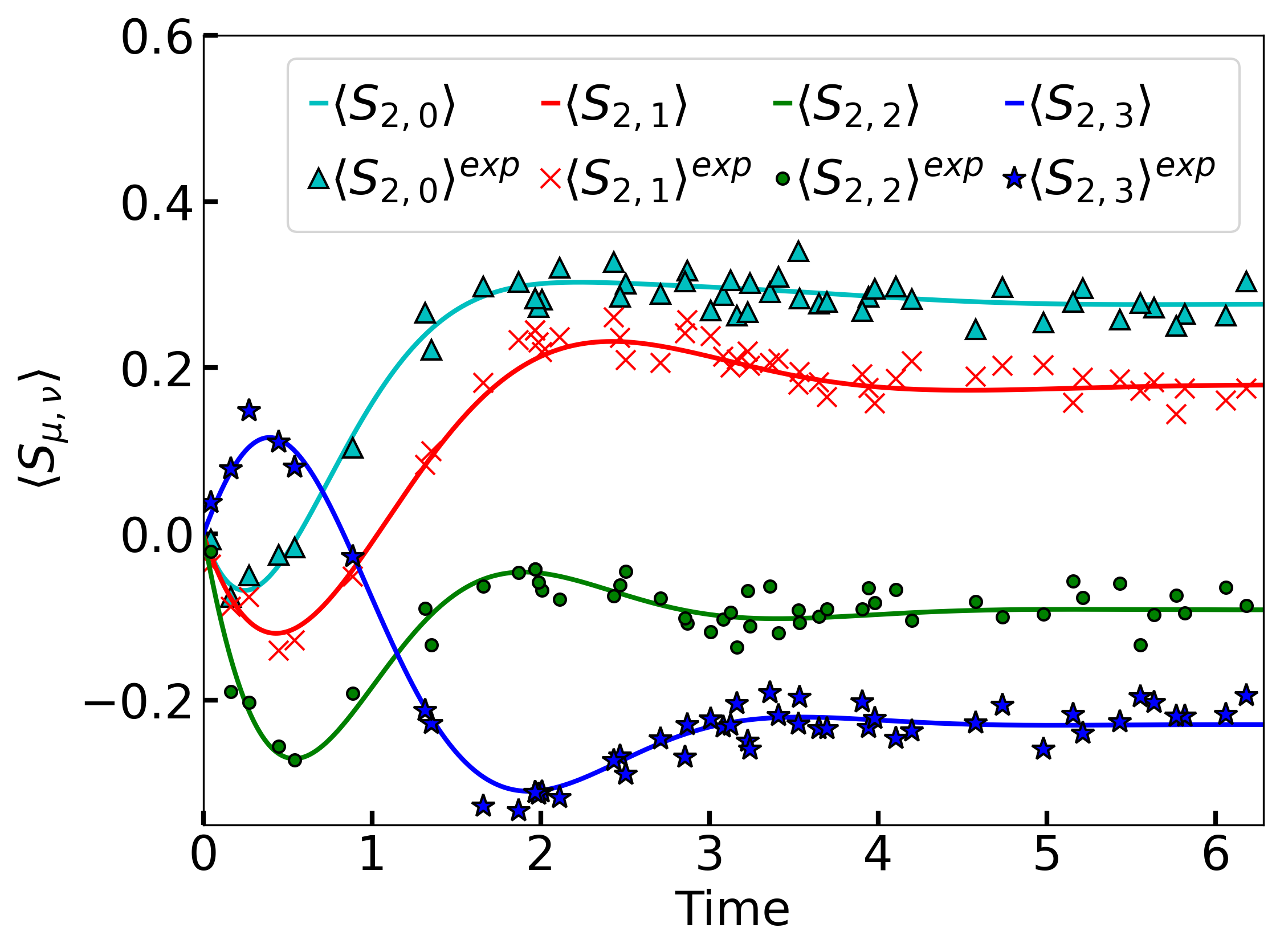}
    \caption{Expectation values as a function of time for the crosstalk case. The horizontal axis shows the time, and the vertical axis shows the expectation values. Solid lines correspond to $\langle S_{2,\nu} \rangle(t_n)$, the predicted values obtained from the parameters inferred by PINNverse . Markers (triangles, crosses, circles, and stars) represent the training data $\langle S_{2,\nu} \rangle^{\mathrm{exp}}(t_n)$, obtained from simulated data with additive noise (standard deviation 0.02), for $\mu = 2$ and $\nu = 1, 2, 3$.Training was performed using $N_c = 50$ collocation points.
    }
    \label{fig:observables_dynamics}
\end{figure}

Figure~\ref{fig:observables_dynamics} displays the time-dependent behavior of the two-qubit observables $\langle S_{\mu,\nu} \rangle(t)$ considering a single run of the PINNverse procedure. Symbols represent the data $\langle S_{\mu,\nu} \rangle^{exp}(t_j)$ considering a Gaussian error with $\sigma=0.02$. The solid lines are the results obtained from the estimated crosstalk Hamiltonian and dissipation parameters using the PINNverse. These results demonstrate a strong agreement between the data and the reconstructed trajectories for most observables, with an MAPE of $0.0023$ ($0.23\%$). This suggests that the inferred model can accurately reproduce the crosstalk parameters even in the presence of decaying components of the non-unitary dynamics.

\begin{figure}
    \centering
    \includegraphics[width=1\linewidth]{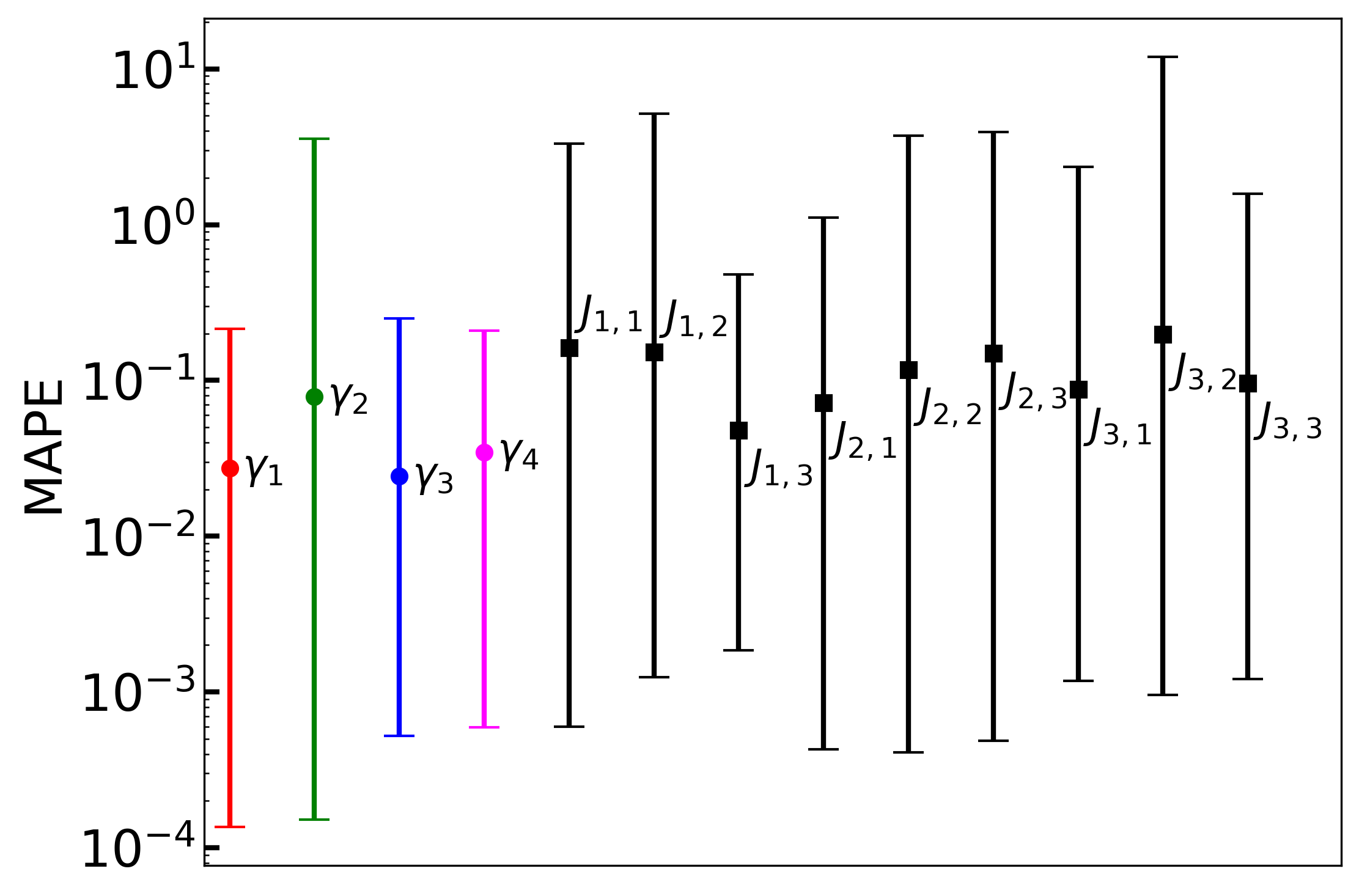}
    \caption{The MAPE plotted on a logarithmic scale for each parameter estimated from data  with Gaussian error considering a standard deviation of 0.02 and \( N_c = 50 \) collocation points. MAPE for the decay rates \( \gamma_1, \gamma_2, \gamma_3, \gamma_4 \) and for crosstalk terms \( J_{\mu,\nu} \) are depicted. Error bars indicate the maximum and minimum values of the MAPE across multiple PINN runs.}
    \label{fig:parameter_errors}
\end{figure}

Figure \ref{fig:parameter_errors} presents the MAPE associated with the estimation of each dissipation parameter $\gamma_i$ and crosstalk coefficients $J_{\mu,\nu}$ in the Hamiltonian, considering fixed values of the noise standard deviation ($\sigma=0.02$) and the number of collocation points ($N_c = 50$). These results indicate that parameters $\gamma_1, \gamma_2, \gamma_3, \gamma_4$ are typically estimated with an error below $1\%$, while crosstalk coefficients $J_{\mu,\nu}$ present the average MAPE below $1.3\%$. These results demonstrate an accurate identification of both dissipation and crosstalk parameters when there is a small Gaussian error and a sufficient number of experimental data.

\begin{figure}[!ht]
    \centering
    \includegraphics[width=1\linewidth]{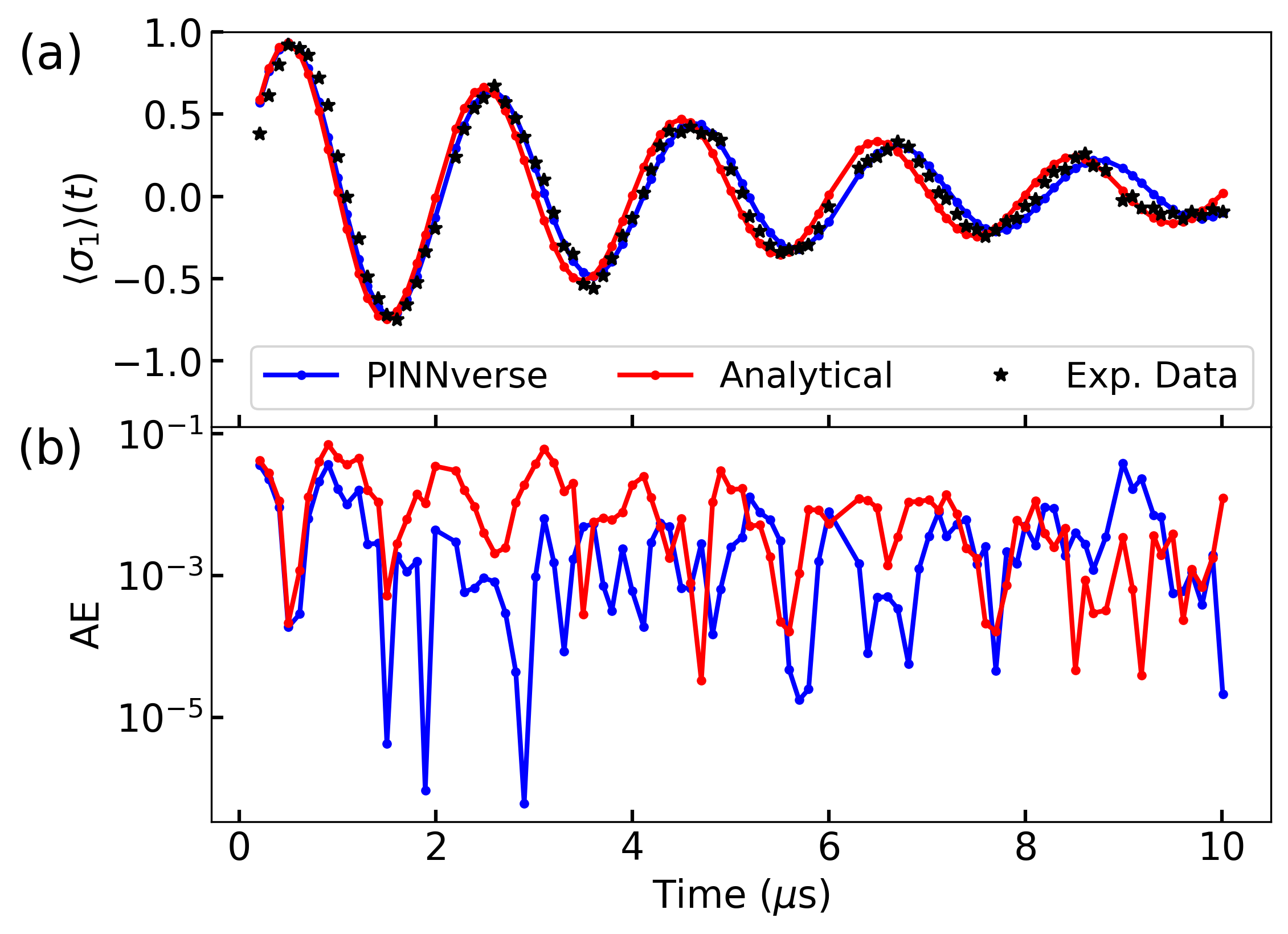}
    \caption{ (a) The observable $ \langle \sigma_1 \rangle(t) $ as a function of time, with experimental data represented by black stars and the curves correspond to the solutions obtained using PINNverse (blue line with circular markers) and the analytical results (red line with circular markers). (b) The absolute error as a function of time for both models, highlighting the comparison between the PINNverse (blue line with circular markers) and the analytical solution (red line with circular markers).}
    \label{fig:1qubitsx}
\end{figure}

\subsection{Experimental data for one qubit.}

In this section, we apply the procedure to a real experiment performed on a single qubit~\cite{Ficheux2018}. The Hamiltonian in Eq.~(\ref{eqTwoqubit}) must be reduced to a one qubit case, therefore we have:
\begin{equation}
    H = \hbar\sum_{\mu=1}^{3} J_{\mu} \, \sigma_\mu,
    \label{eqOnequbit}
\end{equation} 
\noindent where \( J_{\mu} \) characterizes the unitary dynamics of the one-qubit system. The Lindblad operators for this case are
$\mathcal{L}_1 = \sigma_{3}$, $\mathcal{L}_2 = \sigma_{-}$, and $\mathcal{L}_3 = \sigma_{+}$, where $\mathcal{L}_1$ corresponds to the dephasing channel and $\mathcal{L}_2$ and $\mathcal{L}_3$ correspond to the amplitude-damping channel at finite temperature, where these two last dissipative terms describe the decay from the excited to the ground state and vice-versa. In Ref.~\cite{Ficheux2018}, authors assume that only $J_2$ is different from zero, which enables to solve the problem analytically (see Ref.~\cite{Ficheux2018} for more details).
Using PINNverse, we found $J_1 = 2.4\times10^{-2} \text{ MHz}, J_2 =-1.52 \text{ MHz}, \text{and } J_3=-1.08\times10^{-2}\text{ MHz}$, while in Ref.~\cite{Ficheux2018} it was used the following values $J^{\mathrm{exp}}_1=J^{\mathrm{exp}}_3=0$ and $J^{\mathrm{exp}}_2=-1.57 \text{ MHz}$.  For the decay rates, we found $\gamma_1 = 1.26\times10^{-1}\text{ MHz}, \gamma_2 = 7.89\times10^{-2} \text{ MHz},  \text{and } \gamma_3= 4.39\times10^{-5} \text{ MHz}$, while  the experimental results are $ \gamma^{\mathrm{exp}}_1 = 1.28\times10^{-1} \text{ MHz}, \gamma^{\mathrm{exp}}_2 = 6.5\times10^{-2}\text{ MHz}, \text{and } \gamma^{\mathrm{exp}}_3= 1.33\times10^{-4} \text{ MHz}$ . One can notice that there is a good agreement between the most relevant parameters (larger values) obtained by the PINNverse and the ones found in the experimental fitting~\cite{Ficheux2018}. To further probe the efficiency of PINNverse, we show in Fig.~\ref{fig:1qubitsx}(a) the evolution of the observable \( \langle \sigma_1 \rangle(t) \), calculated with the parameters obtained using the PINNverse method (blue line with circular markers) and from the analytical solution (see Ref.~\cite{Ficheux2018}) (red line with circular markers), compared to the experimental data (black stars). Both theoretical solutions show a good agreement to the experimental data, indicating that the PINNverse framework can reproduce the observed behavior of the open quantum system accurately.  

The Absolute Error(AE) at each time  $t_n $ is defined as the absolute difference between the experimental and theoretical expectation values of the observables:
\begin{equation}
	\text{AE}(t_n)=|\langle \sigma_\mu\rangle ^\text{exp}(t_n)-\langle \sigma_\mu\rangle (t_n)|,\label{eq:ae}
\end{equation}
where $ \mu = 1, 2, 3 $. Here, \( \langle \sigma_\mu \rangle^\text{exp}(t_n) \) denotes the experimentally measured expectation value, and \( \langle \sigma_\mu \rangle(t_n) \) corresponds to the theoretical prediction from the model, for each observable at time \( t_n \).
Figure~\ref{fig:1qubitsx}(b) shows the AE as a function of time for both theoretical approaches. One can observe that the PINNverse framework consistently exhibits a lower error rate compared to the analytical model described in the Ref.~\cite{Ficheux2018}. To quantitatively check this difference, we evaluate the Mean Absolute Error (MAE), which is the sum of Eq.~\eqref{eq:ae} divided by the number of instant of times $t_n$. The MAE for PINNverse related to Fig.~\ref{fig:1qubitsx}(b) reaches the value $4.8\times10^{-3}$, while the MAE for the analytical model is $1.1\times10^{-2}$. This result shows that PINNverse is more precise on average to predict the experimental data.

\begin{figure}[!ht]
    \centering
    \includegraphics[width=1\linewidth]{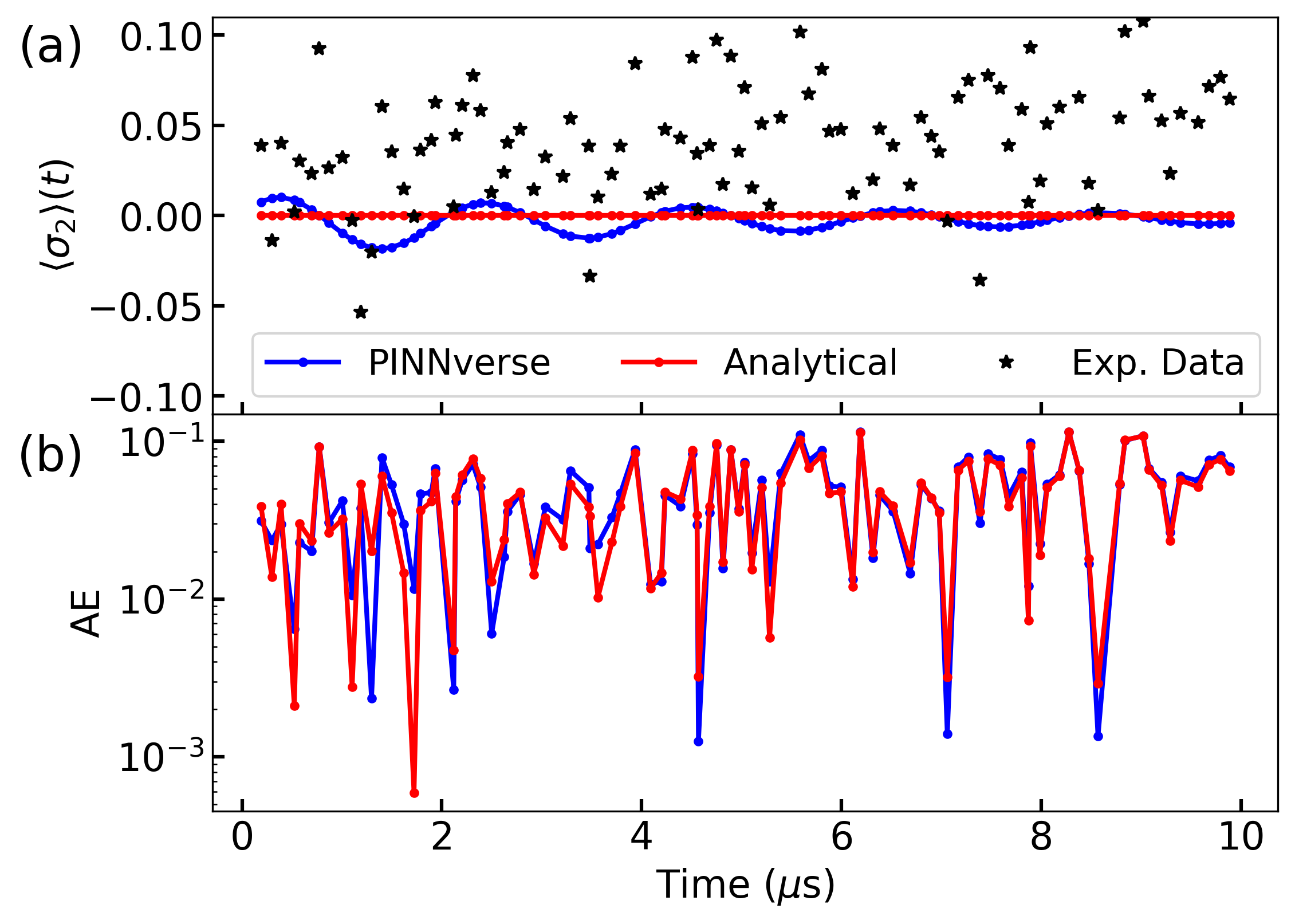}
    \caption{ (a) The observable $ \langle \sigma_2 \rangle(t) $ as a function of time, with experimental data represented by black stars and the curves correspond to the solutions obtained using PINNverse (blue line with circular markers) and the analytical results (red line with circular markers). (b) The absolute error as a function of time for both models, highlighting the comparison between the PINNverse  and the analytical solution .}
    \label{fig:1qubitsy}
\end{figure}

In Fig.~\ref{fig:1qubitsy}(a) we show the temporal evolution of the observable \( \langle \sigma_2\rangle (t)\), where the blue curve with circular markers is obtained with the PINNverse parameters, while the red curve with circular markers uses the null solution of the Ref.~\cite{Ficheux2018}. In this particular case, neither theoretical method fits the experimental data very well because the results are too stochastic and we must remember that our procedure intends to fit the whole set of experimental data composed of $\langle \sigma_\mu\rangle(t) ^\text{exp}$ for $\mu=1,2,$ and 3. The AE shown in Fig.~\ref{fig:1qubitsy}(b) for both theoretical approaches are very similar, as can also be seen by the MAE, which is 0.0469 for the PINNverse and 0.0451 for the analytical solution. 

\begin{figure}[h]
    \centering
    \includegraphics[width=1\linewidth]{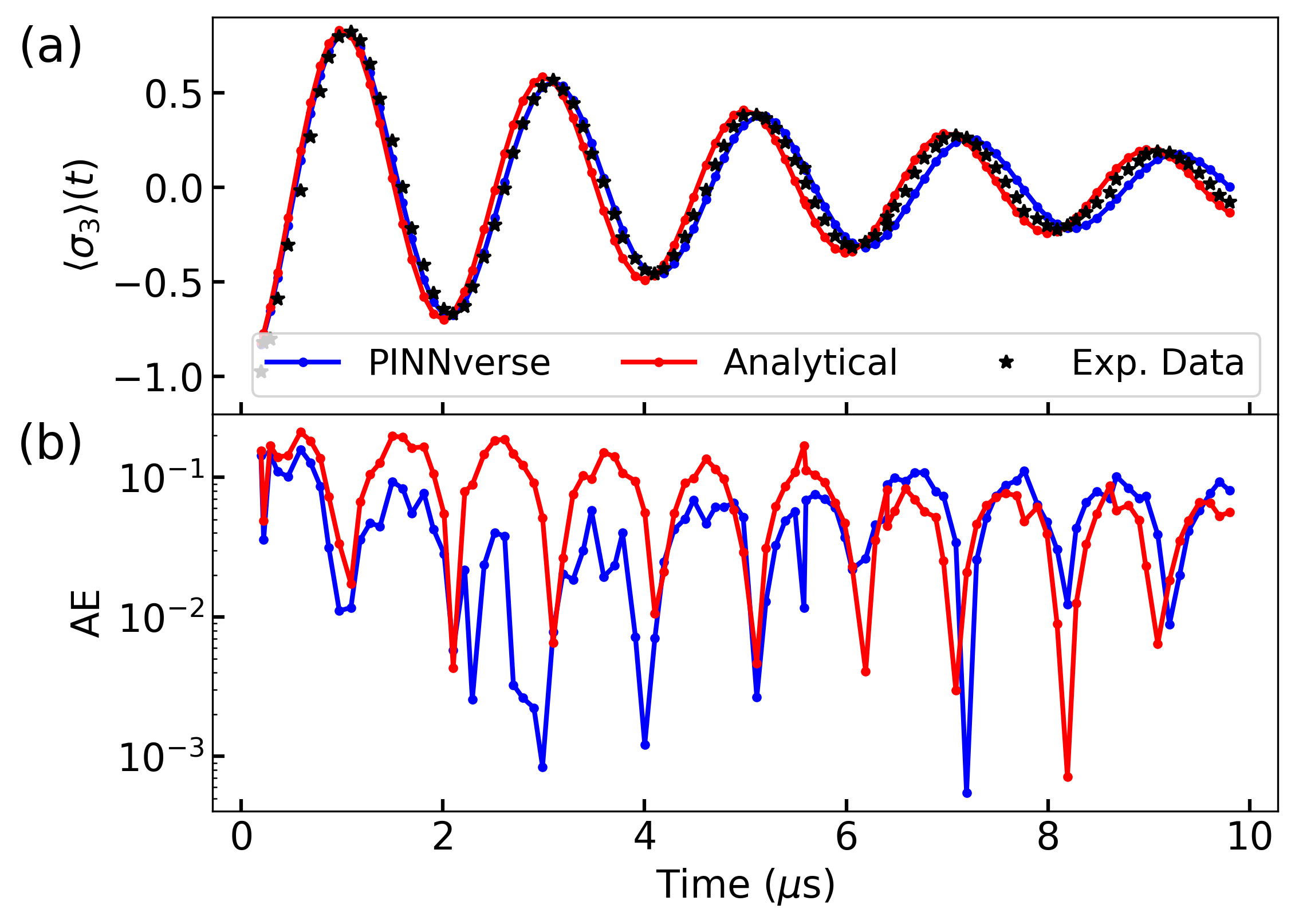}
    \caption{ (a) The observable $ \langle \sigma_3 \rangle(t) $ as a function of time, with experimental data represented by black stars and the curves correspond to the solutions obtained using PINNverse (blue line with circular markers) and the analytical results (red line with circular markers). (b) The absolute error as a function of time for both models, highlighting the comparison between the PINNverse and the analytical solution }
    \label{fig:1qubitsyz}
\end{figure}

Figure~\ref{fig:1qubitsyz}(a) show that both theoretical models show good agreement with the experimental data, although the PINNverse framework  offers a slightly better fit up to the time 6$\mu s$. This can be clearly seen in Fig.~\ref{fig:1qubitsyz}(b), where the absolute error values obtained from the PINNverse are smaller than those found from the analytical model. 
To further check this result, one can see that the MAE obtained for the PINNverse framework is 0.0524, while for the reference model paper is 0.0783. The slight lower MAE obtained using the PINNverse parameters indicates a higher accuracy in fitting the experimental data, when compared to the analytical results.

\section{Conclusions}\label{sec:conclusions}

This work demonstrates that PINNverse offers a powerful tool for characterizing open quantum systems. By explicitly incorporating the Lindblad master equation into the neural network training, the method accurately identifies both Hamiltonian parameters and decaying coefficients. It maintains high performance even in the presence of small stochastic noise in the measurement data. In the case of crosstalk errors characterization, the method successfully reconstructs both the coherent and decoherent dynamics of a two-qubit system. Specifically, the MAPE for the estimated expectation values of two-qubit observables reaches a value of 0.23\% using $N_c = 50$ collocation points, clearly demonstrating the model’s robustness and precision in capturing crosstalk effects. Additionally, using the single-qubit experimental data, PINNverse outperformed the analytical solution, achieving lower mean absolute errors across multiple observables: $4.8 \times 10^{-3}$ for $\langle \sigma_1 \rangle(t)$, $4.69 \times 10^{-2}$ for $\langle \sigma_2 \rangle(t)$, and $5.24 \times 10^{-2}$ for $\langle \sigma_3 \rangle(t)$. These results confirm the method’s strong capability to learn the physical parameters of a real system, where non-unitary effects and errors from experimental preparation and measurement data are present.





\section*{Acknowledgments}
The authors are grateful for financial support from the Brazilian Agencies FAPESP, CNPq, and CAPES. L.K.C. and I.C. thanks the Brazilian Agency FAPESP (Grant Nos. 2024/09298-7 and 2023/04987-6) and also National Institute of Science and Technology for Quantum
Information (CNPq INCT-IQ 465469/2014-0) for supporting this research.
\bibliography{references}

\end{document}